\documentclass[showpacs,aps,prd,nofootinbib,floatfix,amsmath,amssymb]{revtex4}
\usepackage{graphicx}
\begin{document}

\makeatletter
\newbox\slashbox \setbox\slashbox=\hbox{$/$}
\newbox\Slashbox \setbox\Slashbox=\hbox{\large$/$}
\def\pFMslash#1{\setbox\@tempboxa=\hbox{$#1$}
  \@tempdima=0.5\wd\slashbox \advance\@tempdima 0.5\wd\@tempboxa
  \copy\slashbox \kern-\@tempdima \box\@tempboxa}
\def\pFMSlash#1{\setbox\@tempboxa=\hbox{$#1$}
  \@tempdima=0.5\wd\Slashbox \advance\@tempdima 0.5\wd\@tempboxa
  \copy\Slashbox \kern-\@tempdima \box\@tempboxa}
\def\FMslash{\protect\pFMslash}
\def\FMSlash{\protect\pFMSlash}
\def\miss#1{\ifmmode{/\mkern-11mu #1}\else{${/\mkern-11mu #1}$}\fi}
\makeatother

\title{About gauge invariance in compactified extra dimensions}
\author{H. Novales--S\' anchez and J. J. Toscano}
\address{Facultad de Ciencias F\'{\i}sico Matem\'aticas,
Benem\'erita Universidad Aut\'onoma de Puebla, Apartado Postal
1152, Puebla, Puebla, M\'exico.}
\begin{abstract}
Gauge theories formulated in a space--time manifold that includes compact extra dimensions can show a nontrivial gauge structure. Depending on whether the gauge parameters propagate or not in the extra dimensions, two different Kaluza--Klein theories can arise when the extra dimensions are compactified. A comparison between these two possibilities, in the context of a five dimensional theory, is presented from both the theoretical and phenomenological viewpoints. The phenomenological implications of these theories are contrasted by discussing the one--loop decay of the Higgs boson into two photons. It is shown that the amplitude for this decay differs substantially from one approach to the other and that such a difference is intimately related to gauge invariance.
\end{abstract}


\maketitle

Many efforts have been dedicated to extra dimensional Standard Model (SM) extensions since the crucial works by Antoniadis, Arkani-Hamed, Dimopoulos and Dvali~\cite{TeVsz}, where ${\rm TeV}^{-1}$--sized extra dimensions were considered for the first time. In the present paper, we will focus in five dimensional theories in the context of universal extra dimensions~\cite{UED} (UED). The standard four dimensional coordinates will be denoted by $x$, whereas the fifth dimension coordinate will be represented by $y$. Capital roman indices will run over the five space--time coordinates. Gauge and four dimensional Lorentz indices will be denoted by $a,b,c,\ldots$ and $\mu,\nu,\ldots$, respectively, while KK modes indices will be placed between parentheses. The five dimensional theory is given by the following action,
\begin{equation}
S=\int d^4x\, {\cal L}_4\, ,
\end{equation}
with
\begin{equation}
\label{el}
{\cal L}_4=\int dy \, {\cal L}_5\left({\cal F}^a_{MN},F, \Phi, D_M F, D_M \Phi\right)\, ,
\end{equation}
where the extra dimension is assumed to be compactified. In the above expression, ${\cal F}^a_{MN}$ represents the Yang--Mills curvature with gauge fields ${\cal A}^a_M$, whereas $F$ and $\Phi$ are fermion and scalar representations of the gauge group in consideration, and $D_M$ is the covariant derivative. In the literature~\cite{LIT}, it was assumed that the gauge parameters $\alpha^a$, which define the five dimensional ${\rm SU}_5(N)$ gauge transformations, propagate in the fifth dimension, although the role played by the infinite set of KK modes associated with these parameters was not discussed. The four dimensional KK theory  was derived by expanding in Fourier series all the five dimensional fields involved in the theory, including the ${\cal A}^a_M(x,y)$ gauge ones appearing in covariant derivatives. Such an approach does not ruin gauge invariance only if both the five dimensional and the four dimensional theories are governed by the same gauge group~\cite{NT}. Contrastingly, if it is assumed that the gauge parameters propagate in the fifth dimension, only~\cite{NT} gauge covariant objects must be expanded in (\ref{el}). The gauge parameters, $\alpha^a$ determine~\cite{NT} two different four dimensional KK theories depending on whether they are allowed to propagate in the extra dimension or not. The main concern of the present paper is the comparison of such different theories from the theoretical and phenomenological perspectives within the context of the SM in five dimensions. We will discuss the gauge structure of the SM sectors in the context of both formulations and investigate their phenomenological implications by examining their impact on the one--loop decay of the Higgs boson into two photons.

We first restrict our discussion to the five dimensional Yang--Mills sector, whose dynamic variables are vector potentials that can be arranged as $\left({\cal A}_\mu^a(x,y), {\cal A}_5^a(x,y)\right)$. The compactification and integration of the extra dimension produces the so--called zero modes, $A^{(0)a}_\mu$, which are identified as the ordinary Yang-Mills fields in four dimensions. Besides, this procedure generates an infinite set of vector fields, $A^{(m)a}_\mu$, which are the excited modes of the five dimensional gauge fields ${\cal A}_\mu^a$. Finally, this process dissociates the last component of the vector potentials, the ${\cal A}_5^a$ field, and an infinite set of scalars, $A^{(n)a}_5$, also arises. The supposition that the gauge parameters depend on the fifth dimension leads to an infinite set of gauge parameters specifying an infinite number of infinitesimal gauge transformations, which can be divided~\cite{NT} into two types: the ordinary four dimensional Yang--Mills gauge transformations, or {\it standard gauge transformations} (SGT), which are determined by the zero modes of the gauge parameters, $\alpha^{(0)a}$; and another sort of transformations that we called the {\it nonstandard gauge transformations} (NSGT), which are infinite in number and are defined by the excited modes of the gauge parameters, $\alpha^{(m)a}$. In this context, the KK--tower expansions must be performed~\cite{NT} directly on the five dimensional covariant objects, which in this case are the curvatures, as such a process preserves~\cite{NT} gauge invariance after the integration of the fifth dimension coordinate. We call the four dimensional theory so obtained, the {\it pure gauge Kaluza--Klein theory} (GKKT), and denote it by ${\cal L}_{\rm 4YM}$. The degeneracies associated with the SGT and the NSGT can be removed~\cite{NT} by processes that are different and independent of each other. The fixation of the gauge for the GKKT, with respect to the NSGT, is conveniently achieved~\cite{NT} by introducing a SGT--covariant gauge--fixing scheme. The GKKT has been consistently quantized~\cite{NT} and the most general Faddeev--Popov ghost term for the KK excited modes derived~\cite{NT}. It is remarkable that the one--loop effects of the GKKT on any light Green's function are renormalizable~\cite{NT,NT2}, which has been phenomenologically illustrated~\cite{FMNRT}, recently. Such result does not hold, in general, at the two--loop or higher orders, or when two or more extra dimensions are introduced. If the gauge parameters are not allowed to propagate in the extra dimension\footnote{This supposition produces five dimensional Lorentz violation, even before compactification.}, they remain the same at the four dimensional level and determine only~\cite{NT} the SGT. In this case, the objects to KK--expand are~\cite{NT} the gauge fields, instead of the curvatures, and the resulting four dimensional Lagrangian, which we refer to as the {\it non-pure gauge Kaluza--Klein theory} (NGKKT), can be expressed~\cite{NT} as
\begin{equation}
\label{s11}
\hat{\cal L}_{\rm 4YM}={\cal L}_{\rm 4YM}+\Delta {\cal L}\, ,
\end{equation}
with
\begin{equation}
\label{s12}
\Delta {\cal L}=\frac{1}{4}g^2f^{abc}f^{ade}\left(\delta^{rnpq}A^{(n)c}_\nu A^{(q)e\nu}-\delta'^{rnpq}A^{(n)c}_5A^{(q)e}_5 \right)A^{(r)b}_\mu A^{(p)d\mu}\, ,
\end{equation}
where the symbols $\delta^{rnpq}$ and $\delta'^{rnpq}$ are sums of Kronecker delta functions whose precise definitions have been given in Ref.\cite{NT}. The $\hat{\cal L}_{\rm 4YM}$ Lagrangian is invariant under the SGT~\cite{NT}, but not under the NSGT. This approach for the KK--expansions has been the one most commonly followed in the literature~\cite{LIT}, although there has been confusion concerning the gauge structure of the four dimensional theory, as some authors have presented~\cite{KKgt} sets of gauge transformations for all of the KK modes, with an infinite number of gauge parameters defining the variations, which contradicts the above discussion. It has been accentuated~\cite{psegolds} that the scalar fields $A^{(m)a}_5$ are pseudo-Goldstone bosons, but this asseveration is not necessarily true. In the case of the GKKT, the zero--modes $A^{(0)a}_\mu$ and the KK excitations $A^{(m)a}_\mu$ are~\cite{NT} gauge fields under the SGT and the NSGT, respectively, while the scalar fields, $A^{(m)a}_5$, are~\cite{NT} pseudo-Goldstone bosons that can be eliminated~\cite{NT} from the theory. The fields of the NGKKT have a very different nature, as the zero--modes are still gauge fields, but the excited modes $A^{(m)a}_\mu$ are matter fields, i.e., they transform in the adjoint representation of the ${\rm SU}(N)$ group. Even more contrasting is the behavior of the four dimensional scalars, $A^{(m)a}_5$, which are massless scalars that transform in the adjoint representation of ${\rm SU}(N)$ and hence cannot~\cite{NT}  be removed from the theory.

\begin{figure}
\centering\includegraphics[width=6in]{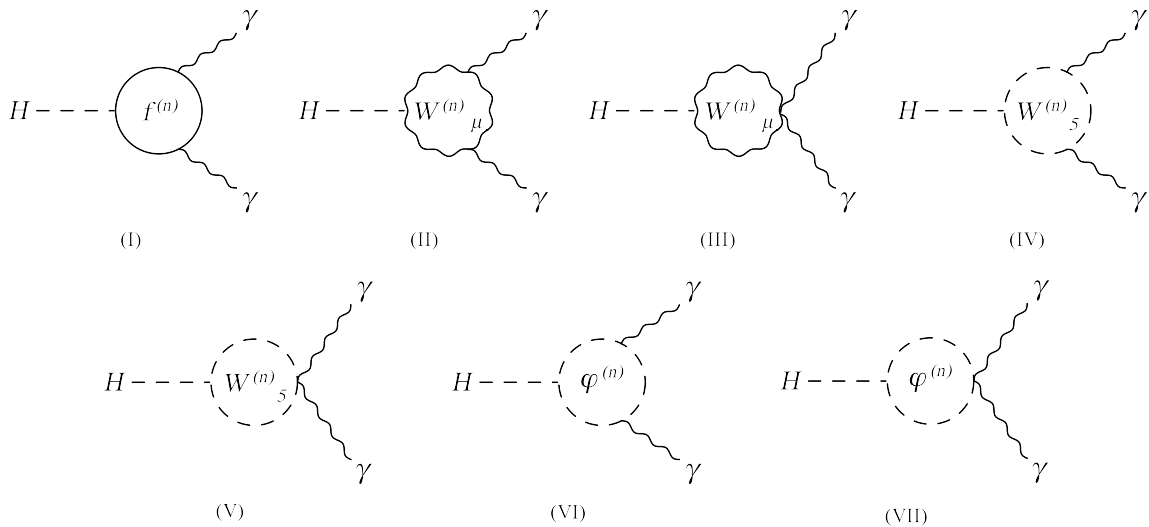}
\caption{\label{Hgg} One--loop order contributions to the vertex $H\gamma\gamma$ from the Yukawa and Higgs five dimensional sectors}
\end{figure}

Phenomenological differences among the GKKT and the NGKKT also occur, and we present an example below. In what follows, the classification of the KK theories into GKKT and NGKKT will be extended to those originated in the whole of the five dimensional SM. From here on, we will use the standard notation for the electroweak gauge fields, $W^a_\mu$ and $B_\mu$. The difference between the Yang--Mills sectors of the GKKT and NGKKT extensions of the SM is given by the $\Delta {\cal L}$ Lagrangian, from which we can identify the tree--level processes   $W^{(m)a}_\alpha W^{(n)b}_\beta\to W^{(m)c}_\mu W^{(n)d}_\nu$ and $W^{(m)a}_5W^{(n)b}_5\to W^{(m)c}_\mu W^{(n)d}_\nu$, which are predictions of the NGKKT SM extension, but not of the GKKT one. Although some different phenomenological implications of these theories are evident from these processes, it would be interesting to show a difference at the level of some electroweak observable. We will consider the decay of the Higgs boson into two photons and use it to qualitatively illustrate that the GKKT and NGKKT are quite different indeed. A wider study on this decay, including other rare processes of the Higgs boson, will be presented elsewhere~\cite{CFNRT}. In the context of UED, the five dimensional sectors of the SM can be classified depending on whether they involve the covariant derivative or not. In the latter category are the Yukawa sector and the Higgs potential, which are respectively given by
\begin{equation}
{\cal L}_{\rm Y}=-\int^{2\pi R}_0 dy \, \left[\lambda_{5u} \bar{Q}(x,y)u(x,y)\tilde{\Phi}(x,y)+ \lambda_{5d} \bar{Q}(x,y)d(x,y)\Phi(x,y)+
\lambda_{5e} \bar{L}(x,y)e(x,y)\Phi(x,y)+h.\, c.\right]\, ,
\end{equation}
\begin{equation}
V=\int^{2\pi R}_0 dy \, \left[\mu^2_5\left(\Phi^\dag(x,y) \Phi(x,y) \right)+\lambda_5\left(\Phi^\dag(x,y) \Phi(x,y) \right)^2\right]\, ,
\end{equation}
where $Q(x,y)$ and $L(x,y)$ are $SU(2)$ doublets of quarks and leptons, respectively, while $\Phi(x,y)$ is the Higgs doublet. In addition, $u(x,y)$ and $d(x,y)$ represent singlets of quarks, whereas $e(x,y)$ stands for a charged lepton singlet. These sectors contribute to the $H\gamma \gamma$ coupling through the $Hf^{(0)}f ^{(0)}$, $Hf^{(m)}f ^{(m)}$, and $H\phi^{(m)-}\phi^{(m)+}$ vertices, with $f^{(0)}$ the standard charged fermions and $f^{(m)}$ ($m=1,2,\cdots $) their excited modes, whereas the charged scalars $\phi^{(m)\pm}$ are excited KK components of the Higgs doublet. Notice that we are using the notation $H\equiv H^{(0)}$. These sectors coincide in both the GKKT and the NGKKT, as they do not involve covariant derivatives, curvatures or covariant objects other than the fermionic fields and the Higgs doublet, so that no differences among the two formulations can be expected in the $H\gamma \gamma$ coupling from these vertices. On the other hand, the coupling of the photon to pairs of charged fermions or scalars is governed by the electromagnetic group, where the GKKT and the NGKKT coincide. So, the fermion (diagram (I) in Fig. \ref{Hgg}) and scalar (diagrams (VI) and (VII) in Fig. \ref{Hgg}) contributions to the $H\gamma \gamma$ coupling coincide in the two theories.

We now turn to discuss the gauge sector contributions. In the GKKT, such contributions are given by the zero $W^{(0)\pm}$ and the excited $W^{(m)\pm}$ modes, which are gauge fields under the SGT and the NSGT, respectively. The scalar fields $W^{(m)\pm}_5$ are the pseudo--Goldstone bosons associated with the gauge fields $W^{(m)\pm}$. Within the scheme of the so--called renormalizable gauges, $R_\xi$, the contributions of the pseudo--Goldstone bosons ($G^{(0)\pm}_W$ and $W^{(m)\pm}_5$) and the ghost (antighost) fields ($C^{(0)\pm}$($\bar{C}^{(0)\pm}$) and $C^{(m)\pm}$($\bar{C}^{(m)\pm}$)) must be considered. However, these contributions are absent in the unitary gauge, in which the contribution to the $H\to \gamma \gamma$ decay is given by the diagrams $(II)$ and $(III)$ in Fig. \ref{Hgg}. On the other hand, in the NGKKT, the excited $W^{(m)\pm}$ modes are not gauge fields, but matter fields, so their contributions to the $H\to \gamma \gamma$ decay must be calculated by using the unitary propagator. Since the difference of the Yang--Mills sectors of these theories is given by the $\Delta {\cal L}$ Lagrangian, which does not involve zero--mode fields, it is clear that the trilinear $W^{(n)\pm}W^{(n)\mp}\gamma$ and quartic $W^{(n)\pm}W^{(n)\mp}\gamma \gamma$ ($n=0,1,\cdots $) vertices induced by the GKKT in the unitary gauge coincide with those generated in the NGKKT. So, these vertices do not produce different contributions to the $H\to \gamma \gamma$ decay in any of the two theories. Another vertex involved in the $W^{(n)\pm}$ contribution to this decay is  $HW^{(n)\pm}W^{(n)\mp}$ ($n=0,1,\cdots $), which is induced by the kinetic term of the Higgs sector. In the GKKT, such term can be written as~\cite{CFNRT}
\begin{equation}
{\cal L}_{\rm 4HK}=(D_\mu\Phi)^{(0)\dag} (x)(D^\mu \Phi)^{(0)}(x)+(D_\mu\Phi)^{(m)\dag} (x)(D^\mu \Phi)^{(m)}(x)+(D_5\Phi)^{(m)\dag} (x)(D^5 \Phi)^{(m)}(x)\, ,\
\end{equation}
where sums over repeated indices, including the modes ones, are assumed. The covariant objects ($(D_\mu\Phi)^{(0)}$, $(D_\mu\Phi)^{(m)}$ and $(D_5\Phi)^{(m)}$) appearing in the above expression can be easily written~\cite{CFNRT} in terms of the Higgs doublet KK excitations ($\Phi^{(0)}$ and $\Phi^{(m)}$), the electroweak fields KK excitations ($W^{(m)a}_\mu$, $W^{(m)a}_5$,$B^{(m)a}_\mu$ and $B^{(m)a}_5$), and a generalization of the covariant derivatives defined in Ref.\cite{NT} to the electroweak gauge group context. The ${\cal L}_{\rm 4HK}$ Lagrangian is~\cite{CFNRT} invariant under both the SGT and the NSGT. On the other hand, when the gauge parameters do not propagate in the fifth dimension, one has the NGKKT. The corresponding Lagrangian can be written as follows:
\begin{equation}
\label{s21}
\hat{{\cal L}}_{\rm 4KH}={\cal L}_{\rm 4KH}+\Delta {\cal L}_{\rm KH}\, ,
\end{equation}
where
\begin{equation}
\label{s22}
\Delta {\cal L}_{\rm KH}=-\Phi^{(n)\dag}\left(\delta^{mnkl}T^{mk}-\delta^{'mnkl}R^{mk} \right)\Phi^{(l)}\, ,
\end{equation}
with
\begin{eqnarray}
T^{mk}&=&g^2W^{(m)a}_\mu W^{(k)b\mu}\frac{\sigma^a}{2}\frac{\sigma^b}{2}+2gg'W^{(m)a}_\mu B^{(k)\mu}\frac{\sigma^a}{2}\frac{Y}{2}+g'^2B^{(m)}_\mu B^{(k)\mu}\left(\frac{Y}{2}\right)^2\, ,\\
R^{mk}&=&g^2W^{(m)a}_5 W^{(k)b}_5\frac{\sigma^a}{2}\frac{\sigma^b}{2}+2gg'W^{(m)a}_5 B^{(k)}_5\frac{\sigma^a}{2}\frac{Y}{2}+g'^2B^{(m)}_5 B^{(k)}_5\left(\frac{Y}{2}\right)^2\, .
\end{eqnarray}
As in the case of the Yang--Mills sector, the GKKT and NGKKT differ in quartic couplings that involve only excited KK modes. This in turn implies that the $HW^{(0)\pm}W^{(0)\mp}$ and $HW^{(m)\pm}W^{(m)\mp}$ couplings coincide in both formulations, which is intimately connected with the renormalizability~\cite{NT,NT2} of the light Green's functions at the one--loop level, as the new physics effects decouple. These vertices are induced by the term $(D_\mu\Phi)^{(0)\dag} (D^\mu \Phi)^{(0)}$ and have the well known SM structure $igm_{W^{(0)}}g_{\mu \nu}$.

Above, we have shown that the contributions of charged fermions and their excited modes, the $W$ gauge boson and their excited modes, and the KK modes of charged scalars associated with the Higgs doublets $\Phi^{(m)}$ are exactly the same in both the GKKT and the NGKKT. However, it is important to notice that the charged scalars $W^{(m)\pm}_5=(W^{(m)1}_5\mp W^{(m)2}_5)/\sqrt{2}$ represent physical degrees of freedom in the NGKKT, so their contribution to the $H\to \gamma \gamma $ decay must be included. Although after the compactification these particles remain massless, they receive mass at the Fermi scale through the vacuum expectation value of $\Phi^{(0)}$. All the excited modes have the same mass $m_{W^{(0)}}$ and the couplings with the Higgs boson are all proportional to $gm_{W^{(0)}}$. The contribution of a charged scalar to the $H\to \gamma \gamma$ decay, which occurs through the diagrams (IV) and (V) of Fig. \ref{Hgg} and is free of ultraviolet divergences, is well known in the literature~\cite{HHG}, so we refrain from presenting it here. For our purposes it is sufficient to establish that the amplitude for the $H\to \gamma \gamma$ decay in the context of the NGKKT can be written as follows:
\begin{equation}
\hat{{\cal M}}(H\to \gamma \gamma)={\cal M}(H\to \gamma \gamma)+\sum_{n=1}^{\infty}{\cal M}^{scalar}(H\to \gamma \gamma)\, ,
\end{equation}
where ${\cal M}(H\to \gamma \gamma)$ is the prediction of the GKKT, which is free of ultraviolet divergences\cite{PSH,CFNRT}. Since ${\cal M}^{scalar}(H\to \gamma \gamma)$ is the same for each mode, it is clear that $\hat{{\cal M}}(H\to \gamma \gamma)$ diverges.

In conclusion, the  two KK gauge theories that arise, depending on whether the gauge parameters propagate in the extra dimension or not, are quite different from both the theoretical and the phenomenological perspectives. The fact that in the GKKT the scalars are pseudo--Goldstone bosons while in the NGKKT they are matter fields is crucial, for in the former the corresponding contributions can be omitted by using the unitary gauge, but in the latter they must be invariably taken into account. Such a behavior is associated to the dynamical variables, which are different in each of these theories. The presence of these relatively light scalar fields in the NGKKT makes this approach less interesting. From the practical point of view, the GKKT is more attractive than the NGKKT because the light Green's functions are renormalizable at the one--loop level, opening the possibility of performing radiative corrections to electroweak observables.

\acknowledgments{ We acknowledge financial support from CONACYT and SNI (M\' exico).}

\end{document}